\documentclass[prl,twocolumn,superscriptaddress]{revtex4-1}
\usepackage[dvipdfmx]{graphicx} %should be commented IN for arxiv
\usepackage{dcolumn}% Align table columns on decimal point
\usepackage{bm}% bold math
\usepackage{color}

\graphicspath{{fig/}}

\begin{document}

\preprint{APS/123-QED}

\title{
Possible High $T_c$ Superconductivity in La$_3$Ni$_2$O$_7$ under High Pressure through  Manifestation of a Nearly-Half-Filled Bilayer Hubbard Model
} 
\author{Hirofumi Sakakibara}
\email{sakakibara@tottori-u.ac.jp}
\affiliation{Advanced Mechanical and Electronic System Research Center(AMES), Faculty of Engineering, Tottori University, 4-10 Koyama-cho, Tottori, Tottori 680-8552, Japan}
\affiliation{Computational Condensed Matter Physics Laboratory, RIKEN, Wako, Saitama 351-0198, Japan}
\author{Naoya Kitamine}
\affiliation{Department of Physics, Osaka University, 1-1 Machikaneyama-cho, Toyonaka, Osaka 560-0043, Japan}
\author{Masayuki Ochi}
\affiliation{Department of Physics, Osaka University, 1-1 Machikaneyama-cho, Toyonaka, Osaka 560-0043, Japan}
\affiliation{Forefront Research Center, Osaka University, 1-1 Machikaneyama-cho, Toyonaka, Osaka 560-0043, Japan}
\author{Kazuhiko Kuroki}
\affiliation{Department of Physics, Osaka University, 1-1 Machikaneyama-cho, Toyonaka, Osaka 560-0043, Japan}

\date{\today}% It is always \today, today,

\begin{abstract}
  Inspired by a recent experiment showing that La$_3$Ni$_2$O$_7$ exhibits high $T_c$ superconductivity under high pressure, we theoretically revisit the possibility of superconductivity in this material.  We find that superconductivity can take place which is somewhat similar to that of the bilayer Hubbard model consisting of the Ni $3d_{3z^2-r^2}$ orbitals. Although the coupling with the  $3d_{x^2-y^2}$ orbitals degrades superconductivity, $T_c$ can still be high enough to understand the experiment thanks to the very high $T_c$ reached in the bilayer Hubbard model.
\end{abstract}

% PACS, the Physics and Astronomy aiueofdsfs
\maketitle
{\it Introduction.}---Seeking for new unconventional high $T_c$ superconductors has been a great challenge ever since the discovery of  the two families of unconventional high-$T_c$ superconductors, cuprates~\cite{CuReview} and iron-based~\cite{FeReview}. Several previous studies have shown that the cuprates are already in an ideal situation in that they are described by a single-orbital Hubbard model near half-filling on a square lattice, and hence their $T_c$ may be difficult to transcend~\cite{Arita3D}.

One possible approach for pursuing even higher $T_c$  is to realize in actual materials the bilayer Hubbard model, for which several studies have shown that the superconducting $T_c$ can be higher than that of the $d$-wave superconducting state in the single-orbital Hubbard model~\cite{KA,MaierScalapino}.  In fact, the bilayer Hubbard model has been widely studied from the past~\cite{Bulut,Scalettar,Hanke,Santos,Mazin,Kancharla,Bouadim,Fabrizio,Zhai,Maier,Nakata,MaierScalapino2,Matsumoto2,DKato,Kainth,Congjun}, and $s\pm$-wave superconductivity~\cite{spmw} is found to be strongly enhanced near half-filling when the vertical electron hopping ($t_\perp$) between the layers is several times larger than the in-plane hopping, and the Fermi level ($E_F$) lies in the vicinity of the edge of one of the bands~\cite{Bulut,KA,Maier,MaierScalapino,Nakata,MaierScalapino2,Matsumoto2,DKato}. 
Nowadays, a band whose edge lies just below or above  $E_F$ is often referred to as an incipient band, and has attracted interest in the study of iron-based superconductors~\cite{DHLee,Hirschfeld,Hirschfeldrev,YBang,YBang2,YBang3,Borisenko,Ding}, bilayer and ladder-type lattices~\cite{Kuroki,MaierScalapino2,Matsumoto,Ogura,OguraDthesis,Matsumoto2,DKato,Sakamoto,Kainth}, and flat band superconductivity~\cite{KobayashiAoki,Misumi,Sayyad,Sayyad2,Aokireview}. 

In fact, one of the present authors proposed that a double layer Ruddlesden-Popper compound La$_3$Ni$_2$O$_7$ can be a good candidate for realizing the bilayer Hubbard model that satisfies the above-mentioned conditions~\cite{Nakata}.  In this material, for which the Ni $3d$ electron configuration is $d^{7.5}$, the $3d_{3z^2-r^2}$ orbitals are elongated in the $z$ (out-of-plane) direction so that $t_\perp$ between the layers is much larger than the in-plane hoppings between the neighboring $d_{3z^2-r^2}$ orbitals, and also the $d_{3z^2-r^2}$ orbitals are nearly half-filled. Hence the $d_{3z^2-r^2}$ portion of the electronic structure appears to be favorable for superconductivity from the above-mentioned viewpoint of the bilayer model,  although deviation from the ideal model arises due to the presence of the Ni $3d_{x^2-y^2}$ bands,  which are nearly quarter-filled, overlapping and hybridizing with the $d_{3z^2-r^2}$ bands.

Given this background, a recent experimental finding that La$_3$Ni$_2$O$_7$ exhibits high $T_c$ superconductivity at high pressures~\cite{MWang}, which in itself has huge impact,  is certainly intriguing. There, it was shown that the material undergoes a superconducting transition with a highest $T_c$ of 80 K under pressure above 14 GPa.
 Already several theoretical studies on this material, which have been performed independently from ours, have appeared right after the discovery of superconductivity~\cite{Luo,Yang,Dagotto,Eremin}.
 In particular, Ref.~\cite{Yang} finds dominance of $s\pm$-wave pairing using functional renormalization group, while Ref.~\cite{Eremin} concludes $d$-wave and $s\pm$-wave being dominant and subdominant, respectively.

Inspired by this experiment, here we theoretically revisit  the possibility of superconductivity in La$_3$Ni$_2$O$_7$ by constructing a four-orbital model that takes into account the crystal structure at high pressures. We find that $s\pm$-pairing superconductivity, which is somewhat similar to that of the bilayer Hubbard model,  can take place with  high $T_c$ that is consistent with the experimental observation. Although the coupling between the $d_{3z^2-r^2}$ and the $d_{x^2-y^2}$ orbitals degrades superconductivity, $T_c$ can still be high because of the very high $T_c$ attained in the bilayer Hubbard model. We also discuss ways to further enhance superconductivity of this material.

\begin{figure}
%\hspace*{-0.5cm}
	\includegraphics[width=9cm]{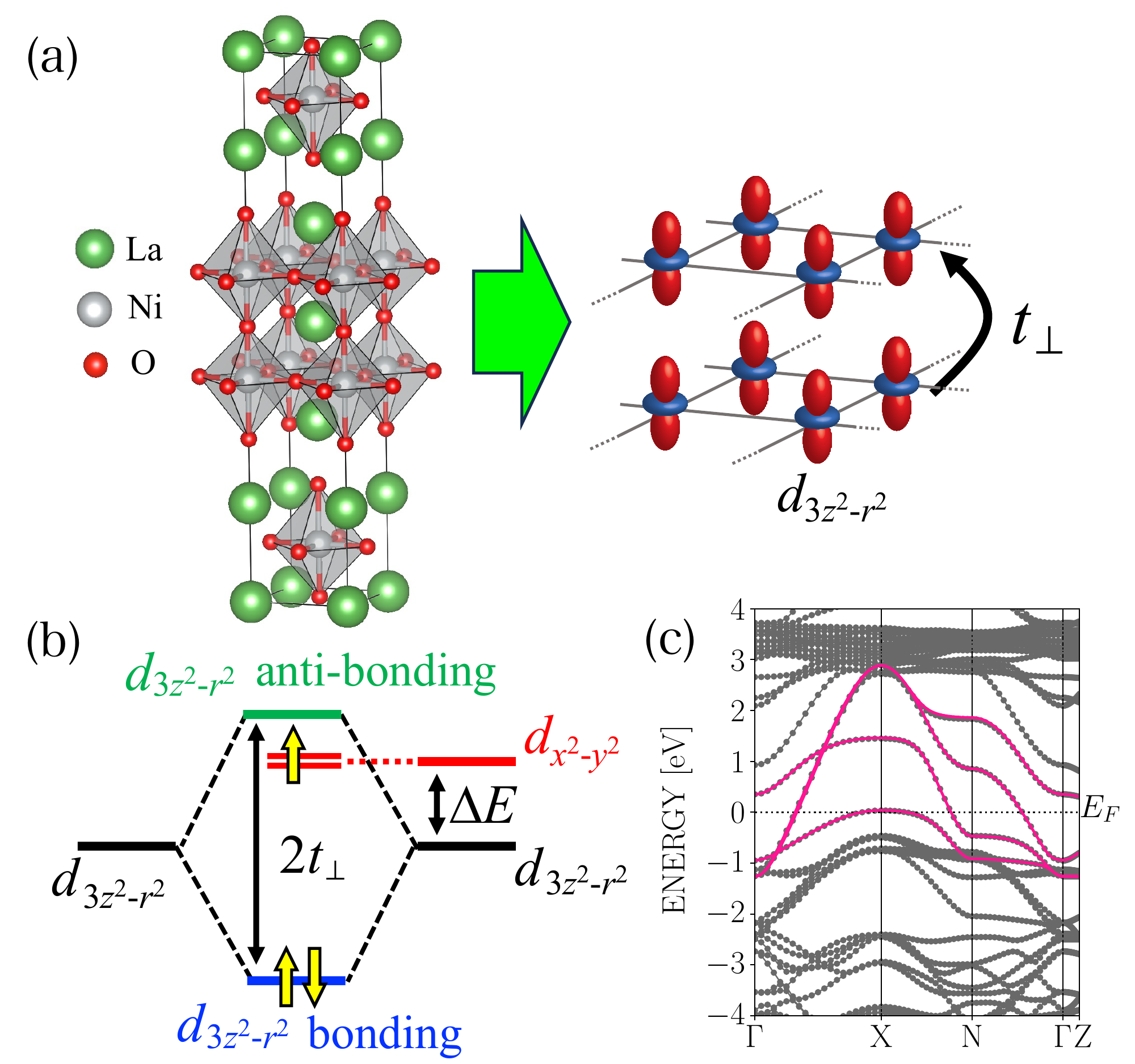}	
	\caption{(a) Crystal structure drawn by VESTA~\cite{VESTA}, (b) the energy diagram 
	of $e_g$ orbitals in our bilayer model, and (c) the first-principles band structure
	of La$_3$Ni$_2$O$_7$ are shown. In the right side of panel (a), schematic figure of the bilayer model of the $d_{3z^2-r^2}$ orbitals is depicted. In panel (c), Wannier-interpolated band structure (pink lines) of 
	the four-orbital model is superposed on the first-principles band structure (gray lines).
	The definition of the energy level offset $\Delta E$ and $t_\perp$, key parameters of this study, are indicated in panel (b) and (a), respectively. 
	}
	\label{fig1}
\end{figure}

{\it Method.}---First, we perform first-principles calculation to obtain the band structure of La$_3$Ni$_2$O$_7$ using the {\footnotesize QUANTUM ESPRESSO} code~\cite{QE}. 
The Perdew-Burke-Ernzerhof parametrization of the generalized gradient approximation (PBE-GGA)~\cite{PBE-GGA}
and the scalar-relativistic version of the optimized norm-conserving Vanderbilt pseudopotentials~\cite{ONCVP} taken from PseudoDojo~\cite{Dojo} are used.
As for the structural parameters, we adopt those given in the extended data table 1 of Ref.~\cite{MWang}  for La$_3$Ni$_2$O$_7$ under the pressure of $P = 29.5$ GPa, namely, the experimentally determined lattice constants as well as the atomic positions determined by theoretical optimization.
Since the orthorhombicity at $P=29.5$ GPa is quite small ($(a-b)/a\sim 1.3 \%$, where $a,b$ are lattice constants for space group $Fmmm$), 
we adopt a body-centered tetragonal structure ($I4/mmm$, Fig. \ref{fig1}(a)) as in La$_2$CuO$_4$, with the in-plane lattice constant determined as an average of the original ones, i.e., $a^*=b^*=(a+b)/2\sqrt{2}$ (see supplemental material~\cite{SM} for details, which includes Refs. \cite{2311.12769,2308.06771,2307.16873,2307.14965,PhysRevB.108.174511,kakoiarxiv}). 
Note that if optimizing not only the internal coordinates but also the lattice parameters within PBE-GGA, the structure indeed goes to $I4/mmm$~\cite{2309.15078,2309.17279}, in agreement with a recent experiment~\cite{2311.09186}.
We take 100 Ry plane-wave cutoff energy, a 12 $\times$ 12 $\times$ 12 $k$-mesh, and an energy width of 0.02 Ry for Gaussian smearing.

We then extract maximally localized Wannier functions~\cite{Marzari,Souza} using the {\footnotesize RESPACK} code~\cite{respack0,respack,respack1,respack2,respack3,respack4,respack5}, 
by which we also obtain the hopping parameters among the Wannier functions. 
We construct a four-orbital model consisting of the $d_{x^2-y^2}$ and the $d_{3z^2-r^2}$ like Wannier orbitals centered at two Ni sites per unit cell.
Important parameter values are given in Table \ref{tab1}.
Figure \ref{fig1}(c) shows superposed band structures given by first-principles calculation and Wannier interpolation, where precise fitting around the Fermi level is achieved.

We explore the possibility of superconductivity %and tendency toward magnetism 
for the obtained low-energy four-orbital model within the fluctuation-exchange (FLEX) approximation~\cite{Bickers,Bickers1991}. 
%\cite{Bickers1989,Bickers1991,Ikeda_omega0,comment7}. 
As the interaction term of the Hamiltonian, we only take the on-site interactions, namely, intraorbital(interorbital) Coulomb interactions $U$($U'$), 
Hund's coupling $J$, and pair hopping $J'$.
We assume the orbital rotational symmetry, namely, we take the same value of $U$ for the $d_{x^2-y^2}$ and the $d_{3z^2-r^2}$ orbitals,
and $U'=U-2J, J=J'$.
Since typical values for cuprate are $U/t=7-10$
 (where $|t|\simeq 0.45$ eV is a typical first-principles value~\cite{Sakakibara1,Sakakibara2,Sakakibara3,Sakakibara4} of the nearest neighbor hopping among the $d_{x^2-y^2}$ orbitals),  we take $U=3$ eV.  
We also take $J=0.1U$, i.e.,  $J=J'=0.3$ eV and $U'=U-2J=2.4$ eV. 
We calculate the self-energy induced by the spin-fluctuation formulated as shown in the literatures~\cite{Lichtenstein,mFLEX1,mFLEX2} in a self-consistent calculation.
The explicit formulae of the irreducible, spin, and charge susceptibilities 
describing the fluctuations are shown in Eqs.(2)-(4) of Ref.~\cite{Sakakibara2}.
The real part of the self-energy at the lowest Matsubara frequency is subtracted in the same manner with Ref.~\cite{Ikeda_omega0}
to maintain the band structure around the Fermi level obtained by first-principles calculation.

The obtained Green's function and the pairing interaction, mediated mainly by spin fluctuations, are plugged into the linearized Eliashberg equation.
Since the eigenvalue $\lambda$ of the linearized Eliashberg equation reaches unity at $T=T_c$, 
we adopt it as a measure of superconductivity at a fixed temperature, $T=0.01$ eV. 
For convenience, we will call the eigenfunction (with the largest eigenvalue) of the linearized Eliashberg equation at the lowest Matsubara frequency $i\omega$(=$i\pi k_{\rm B}T$) the ``superconducting gap function''. We take a 16$\times$16$\times$4 $k$-point mesh and 2048 Matsubara frequencies for the FLEX calculation.
Further details on the model construction and the FLEX calculation are provided in the supplemental materials~\cite{SM}.
%For more details on the FLEX calculation, see the supplemental material\cite{suppl}.

\begin{table}[!h]
\caption{The orbital level offset $\Delta E=E_{x^2-y^2}-E_{3z^2-r^2}$ between the $d_{x^2-y^2}$ and the $d_{3z^2-r^2}$ orbitals, the vertical interlayer hopping $t_{\perp}$ between the $d_{3z^2-r^2}$ orbitals, and
 the nearest-neighbor intralayer hoppings
   $t_{3z^2-r^2}$, $t_{x^2-y^2}$, and $t_{x^2-y^2{\text -}  3z^2-r^2}$ are displayed.
\label{tab1}}
\begin{tabular}{c c c c c c c} \hline\hline
[eV]& \hspace{2pt}$\Delta E$ & \hspace{5pt}$t_{\perp}$ & \hspace{5pt}$t_{3z^2-r^2}$ & \hspace{2pt} $t_{x^2-y^2}$ & \hspace{5pt}$t_{x^2-y^2{\text -} 3z^2-r^2}$ \\\hline
  & \hspace{5pt}$0.372$  \hspace{10pt}& \hspace{5pt}$-0.664$ & \hspace{5pt}$-0.117$ & \hspace{5pt}$-0.491$ &\hspace{5pt} $-0.242$ \\
\hline\hline
\end{tabular}
\end{table} 
 
\begin{figure*}
\begin{minipage}[b]{7.5cm}
\hspace*{-1.5cm}
\includegraphics[width=7.5cm]{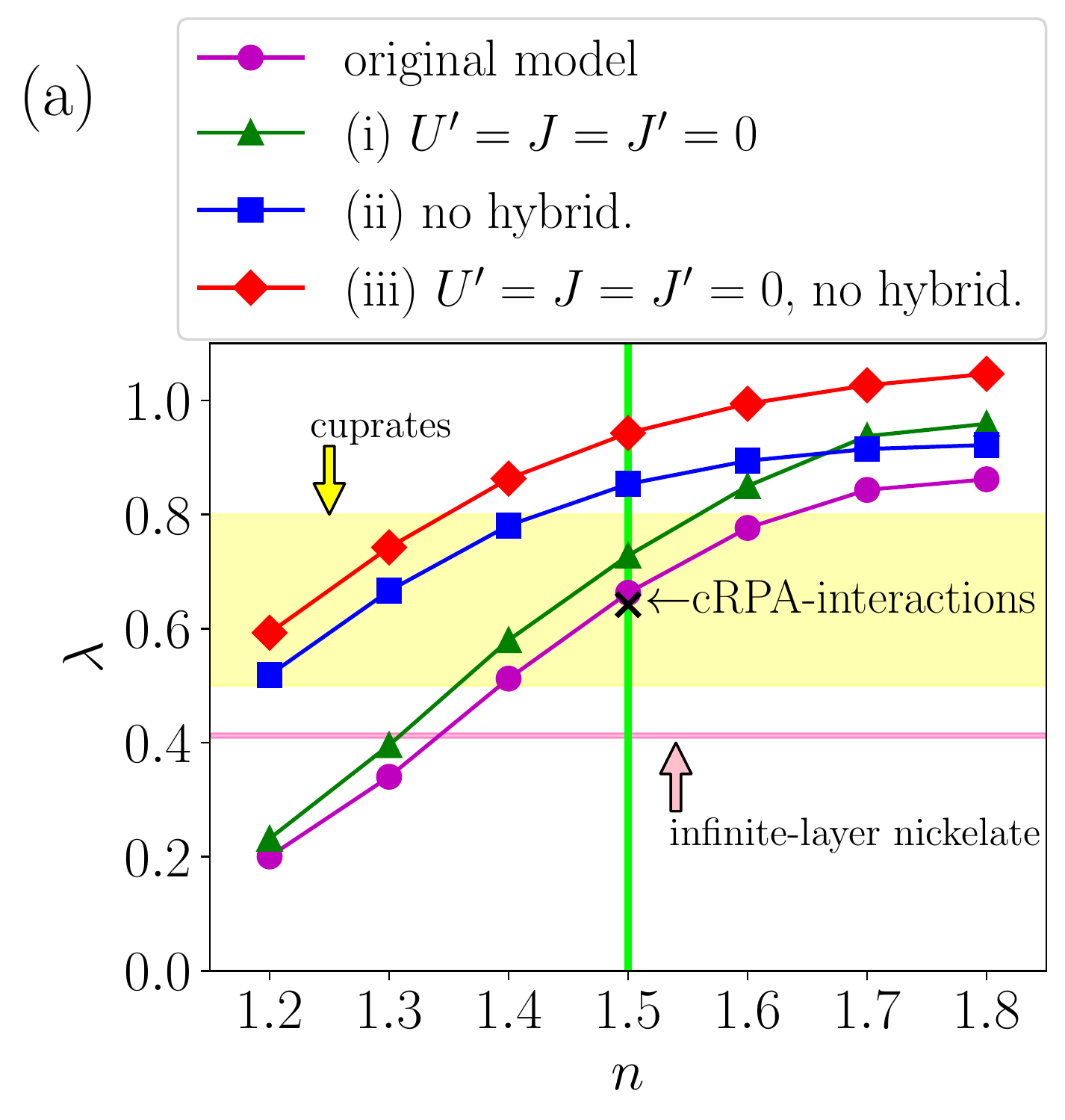}
\end{minipage}
\begin{minipage}[b]{10cm}
\hspace*{-0.5cm}
\includegraphics[width=10.5cm]{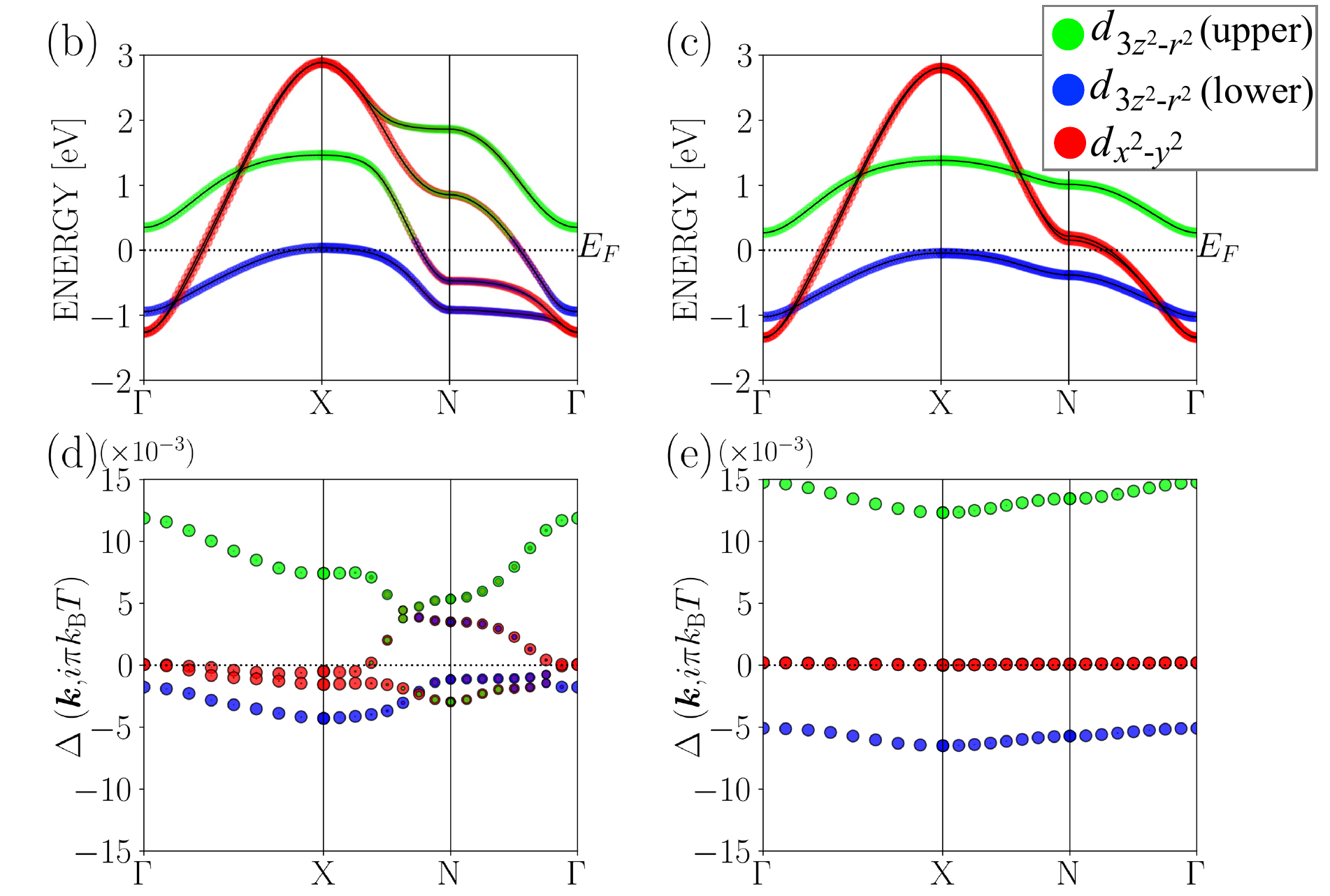}
\end{minipage}
	\caption{(a) The eigenvalue of the Eliashberg equation $\lambda$ at $T=0.01$ eV as a function of the band filling $n$,
	(b)(c) the band structures of the four-orbital model with and without the orbital hybridization for $n=1.5$ (see main text), 
	and (d)(e) the superconducting gap functions of for $n=1.5$ at $T=0.01$ eV in the band representation.
	The inter-orbital interactions are considered in the both calculations for (d) and (e).
	The vertical line in panel (a) corresponds to the stoichiometric composition of La$_3$Ni$_2$O$_7$ ($n=1.5$).
	For this particular band filling, we also plot as a cross symbol the calculation result adopting the interaction parameters evaluated in Ref.~\cite{Werner} using cRPA.
	The yellow (pink) hatched region indicates the range of $\lambda$ calculated for the cuprates~\cite{mrpa2,SakakibaraNi} (infinite-layer nickelate~\cite{SakakibaraNi}) by FLEX at $T=0.01$ eV.
	The strength of the Wannier orbital characters are shown in (b)-(e) with the thickness/radius of the color coded line/circles, where the weight of the $d_{3z^2-r^2}$ orbital is indicated by either green or blue, depending on whether the band energy is above or below $E=0.2$ eV in (b)(d) and $E=0.0$ eV in (c)(e).
	\label{fig2}}	
\end{figure*}

{\it Results and Discussions.}---In Fig. \ref{fig2}(a), we show the eigenvalue of the linearized Eliashberg equation $\lambda$ at $T=0.01$ eV as a function of the band filling $n$, denoted as ``original model''. $n=1.5$ corresponds to the stoichiometric composition of the actual material, and $n$ is varied assuming a rigid band. %In the figure, we also show a yellow shade presenting the range of the typical values of $\lambda$ for the high $T_c$ cuprates obtained in the same way\cite{Sakakibara4,SakakibaraNi}. It can be seen that the present model,  for $n=1.5$ or larger,  exhibits large $\lambda$ values comparable to those of the cuprates, which implies that the calculation results are consistent with the experimental observation of $T_c\sim 80$ K.
In Fig. \ref{fig2}(d), we show the superconducting gap function $\Delta(\bm{k},i\omega)$ of the present model at $n=1.5$ in the band representation. It can be seen that the gap function is large at portions of the band where the $d_{3z^2-r^2}$ orbital component is large, and the bonding and antibonding portions of the $d_{3z^2-r^2}$ bands [see Fig. \ref{fig2}(b)] have opposite signs of the gap. 

In Fig. \ref{fig2}(a), we also show a yellow shade presenting the range of the typical values of $\lambda$ for the high $T_c$ cuprates obtained in the same way~\cite{mrpa2,SakakibaraNi}. It can be seen that the present model,  for $n=1.5$ or larger,  exhibits large $\lambda$ values comparable to those of the cuprates, which implies that the calculation results are consistent with the experimental observation of $T_c\sim 80$ K.
To elaborate on this point, we further perform FLEX calculation at $n=1.5$ adopting more realistic values for the two-body interaction parameters obtained by constrained random phase approximation (cRPA)~\cite{Werner}, and plot the eigenvalue in Fig. \ref{fig2}(a). The obtained value is close to the original value, reinforcing our conclusion. For comparison, we also show the eigenvalue for the infinite layer nickelate obtained in our previous study~\cite{SakakibaraNi}, which falls below the lower bound of the cuprate region and hence also consistent with the experiment~\cite{Hwang}.

\begin{table}[!h]
\caption{Definition of models (i)--(v) (see main text) considered in this Letter. 
Y or N indicates that the effects written in the column headers are present or absent. 
In models (iv) and (v), $\Delta E$ and $t_\perp$ are changed from the original values as listed, respectively.
\label{tab2}}
\begin{tabular}{  c | c c c c} \hline\hline
              & $U',J,J'$ & hybridization & $\delta (\Delta E)$ & $\delta |t_\perp |$ \\\hline
original model & Y& Y & 0 &0  \\%\hline
(i)           & N& Y & 0 &0  \\
(ii)          & Y& N & 0 &0  \\
(iii)         & N& N & 0 &0  \\%\hline
(iv)         & Y& Y & $+0.2$ eV &0  \\
(v)          & Y& Y & 0 & $+0.2$ eV\\\hline\hline
\end{tabular}
\end{table} 

To understand the origin of the large $\lambda$ values,  we study three other models in the same manner, namely, models in which the following couplings between the $d_{3z^2-r^2}$ and the $d_{x^2-y^2}$ orbitals are eliminated:  (i) the interorbital interactions $U'$, $J$, $J'$,  (ii) the hybridization, and (iii) both the interorbital interactions and the hybridization. Definition of the models, including those discussed later, is summarized in Table \ref{tab2}. The band structure without the hybridization is also presented in Fig. \ref{fig2}(c). In model (iii), the $d_{3z^2-r^2}$ and the $d_{x^2-y^2}$ orbitals are completely decoupled, so that the superconducting state is equivalent to that of the bilayer Hubbard model consisting solely of the $d_{3z^2-r^2}$ orbitals. It can be seen that both the hybridization and the interorbital interactions degrade superconductivity of the bilayer Hubbard model, but since $\lambda$ of the bilayer model is significantly large, $\lambda$ of the original model (full model with both the interorbital interactions and the hybridization included) is still large enough to explain the experimental observation. We will further elaborate on the suppression of superconductivity due to inter-orbital coupling in the Supplemental Materials~\cite{SM}.

The nature of the superconducting gap of the original model (Fig. \ref{fig2}(d)) can be more clearly understood by comparing it with that of model (ii) (the model in which the two orbitals are decoupled in one-body level) shown in Fig. \ref{fig2}(e). Here, the gap has opposite signs between the bonding and antibonding $d_{3z^2-r^2}$ bands. It is an $s\pm$-wave superconducting gap in the wide sense of the term in that it changes sign between the two bands, but the antibonding band does not form a Fermi surface.
We stress that even when one of the bands does not intersect the Fermi level, the spin fluctuations with finite energy arise as a pairing glue~\cite{Nakata}. 
Further details on this point are given in the Supplemental Material~\cite{SM}.
The overall resemblance of the superconducting gaps in Figs.~\ref{fig2}(d) and (e) further confirms our picture that the superconductivity in the present model is $d_{3z^2-r^2}$ orbital driven.

In this context, it is also intriguing to give a look into the present system from a strong coupling viewpoint. Calculating within the second order perturbation, the interlayer exchange coupling between the $d_{3z^2-r^2}$ orbitals gives $J_\perp=4t^2_\perp/U\simeq 0.6$ eV for $U=3$ eV, which is quite large compared with, for example, the nearest neighbor superexchange coupling in the cuprates. $J_\perp$ is also much larger than the intralayer hopping between the neighboring $d_{3z^2-r^2}$ orbitals. Such a large $J_\perp$  should lead to opening of a spin gap, and induce interlayer pairing superconductivity~\cite{DagottoScalapino}, whose gap function changes its sign between bonding and antibonding bands in momentum space. This strong coupling picture is indeed consistent with the FLEX results for both the pure bilayer Hubbard model~\cite{Nakata} and the present model. 
%We note that there is no antiferromagnetic {\it ordering} in spin-gapped systems, and this should also apply to the present model of La$_3$Ni$_2$O$_7$. In fact, the Stoner factor of magnetism (the maximum eigenvalue of $U\chi_0(\bm{q},0)$, where $\chi_0(\bm{q},0)$ is the irreducible susceptibility at the lowest Matsubara frequency) at $n=1.5$ is obtained within FLEX as 0.955 for the original model, which is smaller (less tendency toward magnetism) than 0.967 obtained for model (iii), namely, a model that can be considered as equivalent to the bilayer Hubbard model, in which magnetic ordering should not be present.

In Refs.~\cite{Yamazaki,Kitamine}, some of the present authors studied cases where superconductivity {\it emerges or is enhanced} due to the interorbital interactions between the $d_{x^2-y^2}$ and other $d$ orbitals. 
The effect of the interorbital interactions in the present model is the opposite, namely, they degrade superconductivity. 
A large difference is that there is a bonding-antibonding splitting in the $d_{3z^2-r^2}$ bands in the present bilayer system, which might be the reason why the effect of the interorbital interactions is the opposite. Further study on the origin of the difference between the single and bilayer systems is underway.
%and another difference is that in the nickelates considered in Ref. \onlinecite{Kitamine}, the $d_{x^2-y^2}$ band is dilute, 
%while in the present model, the $d_{x^2-y^2}$ band is close to quarter filling.
%The competition between $s\pm$-wave and $d$-wave pairings is expected to be more significant for cases with larger $d_{x^2-y^2}$ band fillings, which leads to a suppression of superconductivity as was discussed
%thoroughly in Refs.\cite{Sakakibara1,Sakakibara2,Sakakibara3,Sakakibara4}.

\begin{figure}
        \hspace*{-0.5cm}
	\includegraphics[width=9.3cm]{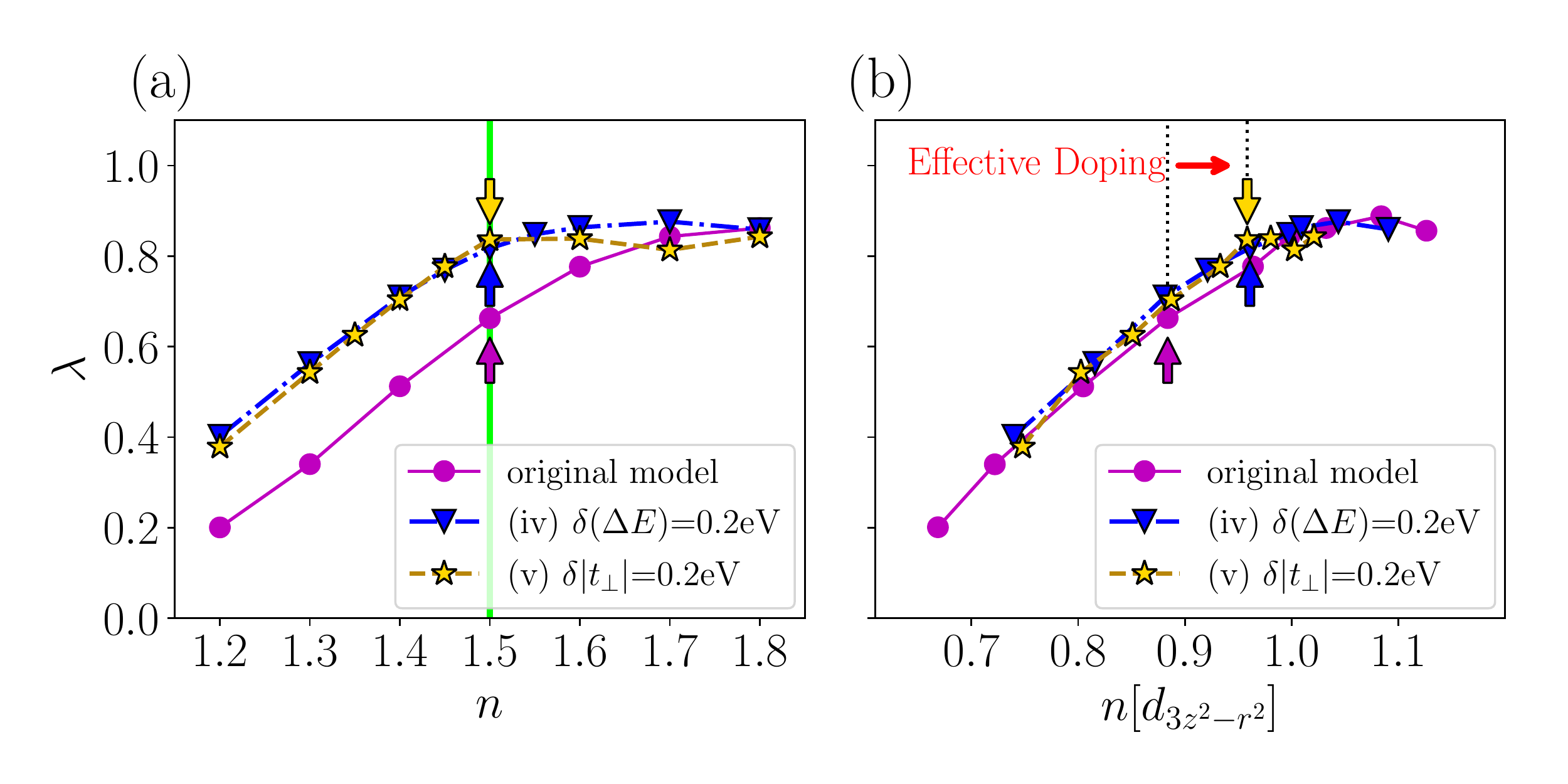}
	\caption{
	(a) The eigenvalue of the linearized Eliashberg equation $\lambda$ at $T=0.01$ eV against the band filling $n$ or (b) the average number of electrons per  
	$d_{3z^2-r^2}$ orbital. The three lines correspond to the three different one-body Hamiltonians:
	the original model and models (iv)--(v) given in Table \ref{tab2}. 
%	the first-principles model,  the model with the orbital level offset $\Delta E$ increased by $\delta(\Delta E)=0.2$ eV,
%	and the model with interlayer hopping $|t_\perp|$ increased by $\delta|t_\perp|=0.2$ eV. 
	The vertical line in panel (a) corresponds to the stoichiometric composition of La$_3$Ni$_2$O$_7$ ($n=1.5$), and
        	the vertical arrows in the both panels indicate data points corresponding to $n=1.5$ for the original model and models (iv)(v).}
	\label{fig3}
\end{figure}

Finally, we discuss possible ways to further enhance superconductivity. The band filling dependence presented in Fig. \ref{fig2} suggests that $T_c$ may be enhanced by doping electrons. In case it is difficult to dope electrons in the actual material, here we propose alternative ways for achieving a similar effect. We consider a model in which (iv) the level offset between the $d_{x^2-y^2}$  and the $d_{3z^2-r^2}$ orbitals $\Delta E=E_{x^2-y^2}-E_{3z^2-r^2}$  is increased by $\delta(\Delta E)=0.2$ eV or (v) $|t_\perp|$  is increased by $\delta|t_\perp|=0.2$ eV (see also Table \ref{tab2}). As depicted in Fig. \ref{fig3}(a), in both models, the band filling dependence of $\lambda$ appears to be shifted toward the left (i.e., toward the smaller $n$ regime), so that larger values of $\lambda$ are attained at $n=1.5$, i.e., the stoichiometric band filling. From a material designing viewpoint, increasing $\Delta E$ and/or $|t_\perp|$ might be achieved by considering mixed anion materials.

The effect of increasing $\Delta E$ and/or $|t_\perp|$ can be understood by counting the number of electrons occupying the $d_{3z^2-r^2}$ orbitals (namely, summing up the $d_{3z^2-r^2}$ orbital weight assuming the non-interacting band structure) for each case. In Fig. \ref{fig3}(b), we plot $\lambda$ against $n$[$d_{3z^2-r^2}$], which is the average number of electrons per $d_{3z^2-r^2}$ orbital. It can be seen that $\lambda$ is mainly determined by $n$[$d_{3z^2-r^2}$] within these three models, which once again supports the picture that the present superconductivity is $d_{3z^2-r^2}$ orbital driven. Here, increasing $\Delta E$ and/or $|t_\perp|$ results in self-doping of electrons from the $d_{x^2-y^2}$ to the $d_{3z^2-r^2}$ orbitals (see Fig. \ref{fig1}(b)). Superconductivity is enhanced as $n$[$d_{3z^2-r^2}$] approaches unity, that is, as the $d_{3z^2-r^2}$ orbital approaches half-filling, so that the electron correlation effects are enhanced, and at the same time, the Fermi level approaches both the bonding band top and the anti-bonding band bottom, thereby shifting the spin fluctuations toward the lower energy regime and making them more effective as a pairing glue.

{\it Summary.}---To summarize, we have studied the possibility of  superconductivity in La$_3$Ni$_2$O$_7$ taking into account the crystal structure under high pressure. The system can be considered as a bilayer Hubbard model of the $d_{3z^2-r^2}$ orbitals coupled with the $d_{x^2-y^2}$ orbitals through interorbital interactions and hybridization. Although the interorbital couplings degrade superconductivity, the $T_c$ can still be high enough to explain the experimental observation, thanks to the very high $T_c$ reached in the bilayer Hubbard model.  We have also discussed possible ways to enhance the superconductivity. Electron doping is likely to enhance superconductivity, but in case this is not feasible, increasing $\Delta E$ and/or $|t_\perp|$ are alternative ways of achieving a similar effect. This is because these modifications result in a self-doping of electrons from the $d_{x^2-y^2}$ to the $d_{3z^2-r^2}$ orbitals.
Studies on material designing along this line is underway.

\begin{acknowledgments}
We are supported by JSPS KAKENHI Grant No. JP22K03512 (H. S.) and JP22K04907 (K. K.).
The computing resource is supported by 
the supercomputer system HOKUSAI in RIKEN,
the supercomputer system (system-B) in the Institute for Solid State Physics, the University of Tokyo, 
and the supercomputer of Academic Center for Computing and Media Studies (ACCMS), Kyoto University.
\end{acknowledgments}

\bibliography{la3ni2o7}

\begin{thebibliography}{86}
\bibitem [{spm()}]{spmw}%
  \BibitemOpen
  \href@noop {} {}\bibinfo {note} {Here, ``$s\pm$-wave'' means that the sign of
  the superconducting gap function is the opposite between the hole-like
  bonding and the electron-like antibonding band of the bilayer model, as shown
  in the right side of Fig. 1 in Ref.~\cite{Kitamine}.}\BibitemShut {Stop}%
\end{thebibliography}

\end{document}